\documentclass{elsart5p}

\usepackage{graphics}
\usepackage{graphicx}
\usepackage{amssymb}

\begin{document}

\begin{frontmatter}



\title{Pressure-induced metal-insulator transition in MgV$_2$O$_4$}
%

\author[usc,iit]{D. Baldomir\corauthref{dbaldo}},
\ead{fadbal@usc.es}
\author[usc,iit]{V.Pardo},
\author[qui]{S. Blanco-Canosa},
\author[qui]{F. Rivadulla},
\author[usc]{J. Rivas},
\author[usc,iit]{A. Pi\~neiro},
\author[iit]{J.E. Arias}

\address[usc]{Departamento de F\'{\i}sica Aplicada, Facultad de F\'{\i}sica, Universidad
de Santiago de Compostela, E-15782 Campus Sur s/n, Santiago de Compostela,
Spain}
\address[iit]{Instituto de Investigaciones Tecnol\'ogicas, Universidad de Santiago de
Compostela, E-15782, Santiago de Compostela, Spain}
\address[qui]{Departamento de Qu\'{\i}mica F\'isica, Facultad de Qu\'{\i}mica, Universidad
de Santiago de Compostela, E-15782, Santiago de Compostela,
Spain}

\corauth[dbaldo]{Corresponding author. Tel: (+34) 981563100 (13960) fax: (+34)
981520676}

\begin{abstract}
On the basis of experimental thermoelectric power results and ab initio
calculations, we propose that a metal-insulator transition takes place at
high pressure (approximately 6 GPa) in MgV$_2$O$_4$.
\end{abstract}

\begin{keyword}
Spinels; first-principles calculations; metal-insulator transition
\PACS 71.30.+h; 71.20.-b; 71.27.+a
\end{keyword}

\end{frontmatter}



Transition metal oxides often present an interplay between different
degrees of freedom (orbital, spin and charge) resulting in the formation of
different superstructures 
\cite{khomskii}. 
Among these, oxide spinels with chemical formula AB$_2$O$_4$,
that present a frustrated pyrochlore lattice for the B cations (transition
metal), are
particularly interesting because they allow to study the importance of
direct cation-cation interactions in a series with single-valent B-B
interactions.

\begin{figure}
\begin{center}
\includegraphics[angle=0,width=0.5\textwidth]{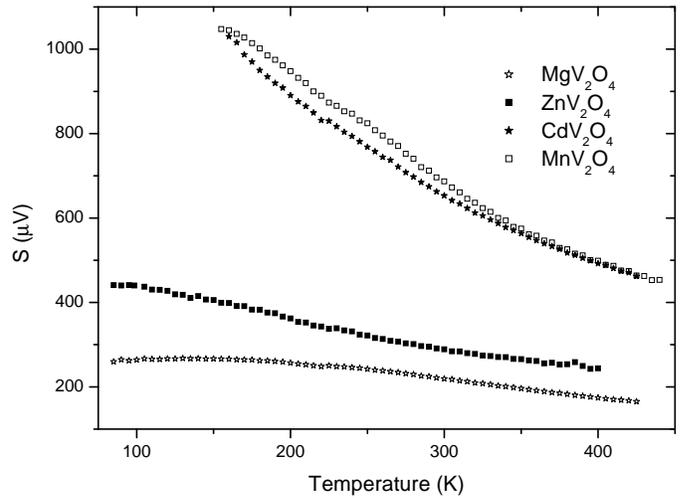}
\end{center}
\caption{Thermoelectric power of the series AV$_2$O$_4$.}\label{fran}
\end{figure}

The series AV$_2$O$_4$ (with A being Cd, Mn, Zn, Mg) approaches a
metal-insulator transition when the V-V distance is reduced sufficiently.
This can be observed in Fig. \ref{fran}.
A reduction in the  
absolute value of the thermoelectric power is evident in Zn and Mg samples, with respect to
Mn and Cd samples. Moreover, typical activated behaviour is no longer  
present in Zn and Mg, and thermopower tends to a constant value at low  
temperature.
This result is fully consistent with a scenario in which  
progressive electronic delocalization in cation-cation bonds occurs as  
the metal-metal distance is reduced across the series. These  
conclusions about partial electronic delocalization are also supported  
by the strong reduction in the magnetic moment observed from CdV$_2$O$_4$ to  
ZnV$_2$O$_4$, in spite of the constant V$^{3+}$\cite{struct}.

The V-V distance, determined from a Rietveld refinement of the X-ray
patterns, reduces along the series from 3.07 \AA\ in the Cd compound
to 2.97 \AA\ for MgV$_2$O$_4$. On the basis of experimental
considerations, 
the critical distance for the
metal-insulator transition was estimated at about 2.94 \AA\ \cite{good}.
One way to study this transition could be to apply pressure to the Mg
compound since it is closest to the critical distance. An experimentally
accessible pressure is expected to produce the transition.

For analyzing this, we have carried out density functional theory calculations on the compound
MgV$_2$O$_4$ within a full-potential, all-electron approach by using
the WIEN2k software\cite{wien}. Electronic correlations were taken into account by
means of the LDA+U approximation\cite{ldau}.

We studied the cubic phase, with lattice parameter $a$= 8.4378 \AA\ measured at room
temperature. The atomic positions are: Mn (0.125, 0.125, 0.125), Cr (0.5,
0.5, 0.5) and O (0.260, 0.260, 0.260) in the space group Fd3m.

\begin{figure}
\begin{center}
\includegraphics[angle=0,width=0.5\textwidth]{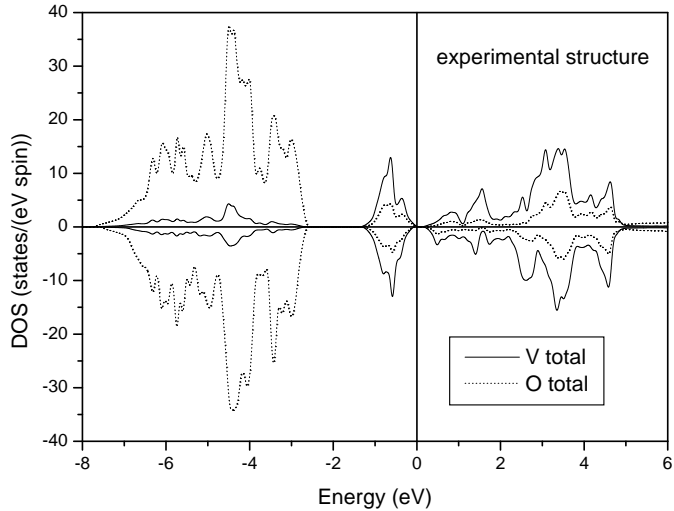}
\end{center}
\caption{DOS plots for all the V and O levels in the unit cell for the
experimental structure. The d-d
character of the narrow gap of only 0.2 eV can be noticed. Upper (lower)
panels show the spin-up (down) channel.}\label{ins}
\end{figure}

\begin{figure}
\begin{center}
\includegraphics[angle=0,width=0.5\textwidth]{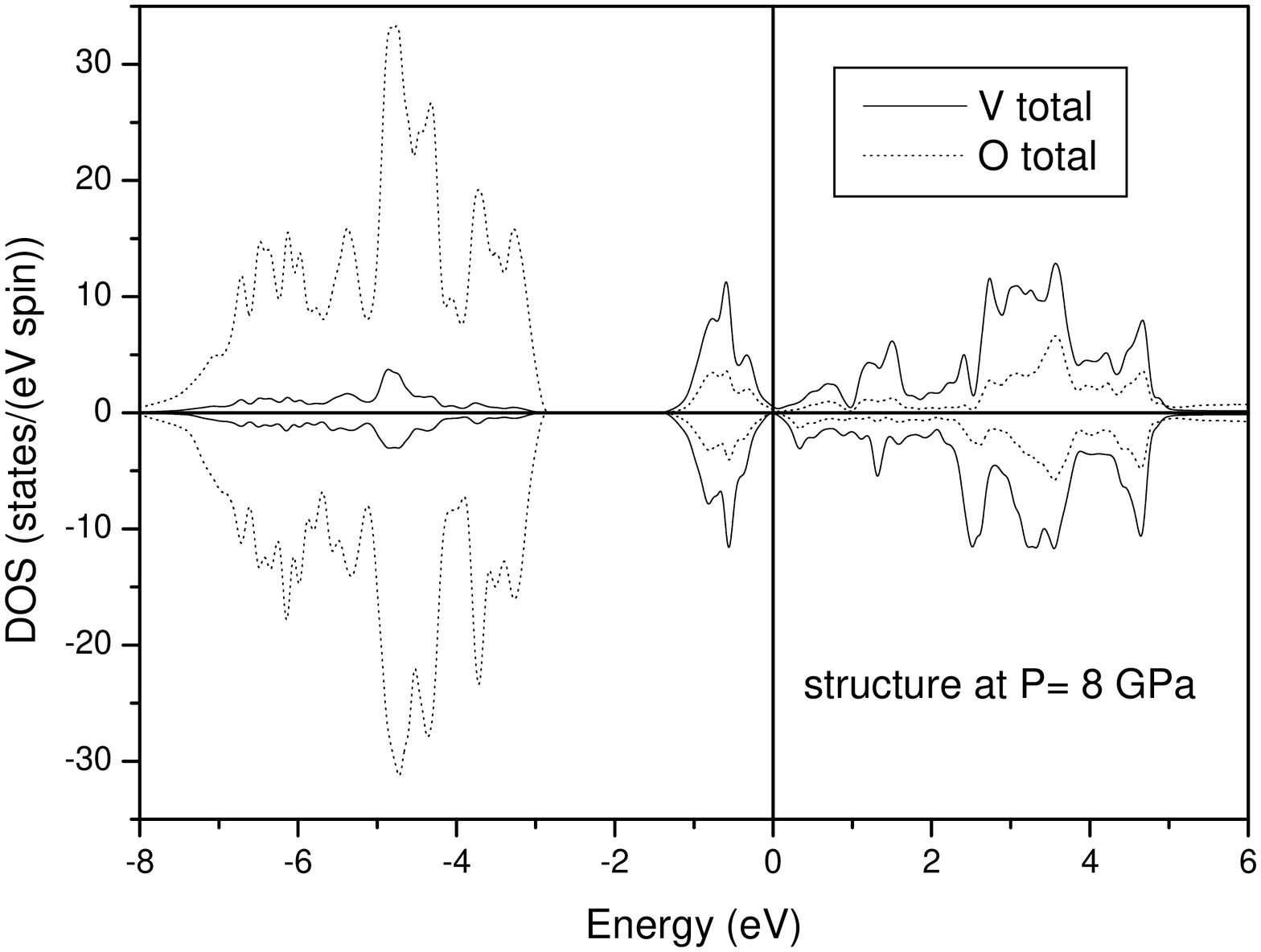}
\end{center}
\caption{DOS plots for all the V and O levels in the unit cell for an
applied pressure of 8 GPa. Observe that metalicity occurs by overlapping
of the d bands that come together due to the small V-V distance. Upper
(lower) panels show the spin-up (down) channel.}\label{met}
\end{figure}

We performed total energy calculations to optimize the volume of the
material within the generalized gradient approximation (GGA) in the
so-called Perdew-Burke-Ernzerhof scheme\cite{gga}. The ground state volume
obtained by this procedure is in agreement with the
experimental one (0.3 \% smaller).
The value of the bulk modulus of 196 GPa is also consistent with
previous results on oxide spinels\cite{prb_oviedo}. It is well known that the
trigonal distortion of the oxygen octahedron surrounding the V$^{3+}$
cations varies with the V-V distance, and this changes slightly
the oxygen positions. We made the approximation of keeping fixed the
oxygen coordinates and only varying the cell parameters.

From our total energy vs. volume calculations, the predicted critical V-V distance (2.94 \AA) is reached at about
5 GPa. For the equation of state around the minimum, we used fittings to
a Murnaghan\cite{murn} and a Birch-Murnaghan equation\cite{birch}, leading both to the same
quantitative results for equilibrium volumes and bulk moduli.
According to our calculations, the transition occurs very close to
that distance, at some 6.5 GPa (V-V distance of 2.937 \AA, in close
agreement with the predicted result).

In Figs. \ref{ins} and \ref{met}, the transition from a narrow-band
insulator to a metal can be observed. For the experimental structure (Fig.
\ref{ins}), the
material is an insulator with a band gap of approximately 0.2 eV. The
value of U was chosen to fit the experimental value of the band
gap\cite{good} (it
turns out to be about 3.1 eV within the so-called self-interaction
corrected LDA+U method\cite{sic}). The material shows the typical
electronic structure of a Mott-Hubbard insulator with a d-d gap. Fig.
\ref{ins} shows the total value of O and V total (including multiplicities) density of states inside the
unit cell. Both the conduction and the valence band have a strong d
character, being the O p contribution to those bands smaller, but
non-negligible.
Fig. \ref{met} shows the density of
states for the structure under an applied pressure of approximately 8 GPa.
The reduced V-V distance leads to a stronger d-d interaction. At a high
enough pressure, the valence and conduction bands
eventually overlap yielding a metallic behaviour.

In summary, a metal-insulator transition is predicted along the series
AV$_2$O$_4$, and should be
obtained by applying pressure to the MgV$_2$O$_4$ compound.
Experimentally, the tendency to metalicity can be observed by analyzing the
variation in the
Seebeck coefficients of the series as the V-V distance gets reduced. Ab
initio calculations predict a critical pressure of some 6.5 GPa needed to
produce the metal-insulator transition.

This work was supported by the Ministry of Science and Education of Spain
(MEC)
through the project MAT2006-10027 and the Xunta de Galicia (Project No. PGIDIT06PXIB209019PR). We are thankful to the CESGA (Centro de
Supercomputaci\'on de Galicia) for the computing facilities. F.R. thanks
the
MEC for support under the program Ram\'on y Cajal.

\end{document}